\documentclass[twocolumn,superscriptaddress,showpacs]{revtex4}
\usepackage{epsf}

\def \mbf#1{{\mbox{\boldmath$#1$}}} % bold greek

\begin{document}

\title{Selective truncations of an optical state using projection synthesis}

\author{Adam Miranowicz}

\affiliation{Institute of Physics, Adam Mickiewicz University,
61-614 Pozna\'{n}, Poland}

\affiliation{SORST Research Team for Interacting Carrier
Electronics, 4-1-8 Honmachi, Kawaguchi, Saitama 331-0012, Japan}

\affiliation{Graduate School of Engineering Science, Osaka
University, Toyonaka, Osaka 560-8531, Japan}

\author{\c{S}ahin Kaya \"Ozdemir}

\affiliation{SORST Research Team for Interacting Carrier
Electronics, 4-1-8 Honmachi, Kawaguchi, Saitama 331-0012, Japan}

\affiliation{Graduate School of Engineering Science, Osaka
University, Toyonaka, Osaka 560-8531, Japan}

\affiliation{CREST Research Team for Photonic Quantum Information,
4-1-8 Honmachi, Kawaguchi, Saitama 331-0012, Japan}

\author{Ji\v r\'\i\ Bajer}
\affiliation{Department of Optics, Palack\'{y} University, 772~00
Olomouc, Czech Republic}

\author{Masato Koashi}

\affiliation{SORST Research Team for Interacting Carrier
Electronics, 4-1-8 Honmachi, Kawaguchi, Saitama 331-0012, Japan}

\affiliation{Graduate School of Engineering Science, Osaka
University, Toyonaka, Osaka 560-8531, Japan}

\affiliation{CREST Research Team for Photonic Quantum Information,
4-1-8 Honmachi, Kawaguchi, Saitama 331-0012, Japan}

\author{Nobuyuki Imoto}

\affiliation{SORST Research Team for Interacting Carrier
Electronics, 4-1-8 Honmachi, Kawaguchi, Saitama 331-0012, Japan}

\affiliation{Graduate School of Engineering Science, Osaka
University, Toyonaka, Osaka 560-8531, Japan}

\affiliation{CREST Research Team for Photonic Quantum Information,
4-1-8 Honmachi, Kawaguchi, Saitama 331-0012, Japan}

\vspace{5mm}

\begin{abstract}
Selective truncation of Fock-state expansion of an optical field
can be achieved using projection synthesis. The process removes
predetermined Fock states from the input field by conditional
measurement and teleportation. We present a scheme based on
multiport interferometry to perform projection synthesis. This
scheme can be used both as a generalized quantum scissors device,
which filters out Fock states with photon numbers higher than a
predetermined value, and also as a quantum punching device, which
selectively removes specific Fock states making holes in the
Fock-state expansion of the input field.

\end{abstract}


\maketitle

%------------------------------------------------------------------
\section{Introduction}

Recent theoretical and experimental works have prompted increasing
interest in quantum state engineering using linear optics. It has
been shown that linear optics can be used for efficient quantum
computation, entanglement manipulation and generation of
nonclassical optical states
\cite{Brien,Yamamoto,Pan,Knill,Book,JMO}. Linear-optical schemes
require single-photon generation and detection, beam splitters
(BSs) and phase shifters (PSs). Parametric down-conversion process
is exploited to build triggered single-photon source, and
avalanche photodiodes are used as photon counters to discriminate
between the absence and presence of photons. Therefore, such
schemes are experimentally realizable with the present level of
optics technology.

In this paper, we study a linear-optical scheme for quantum state
engineering using projection synthesis \cite{Pegg1,Pegg2}. Our
main interest is to employ the scheme to perform the following
transformation
\begin{equation}\label{N01}
|\psi\rangle=\sum_{n=0}^\infty
\gamma_n|n\rangle\longrightarrow|\phi^{(d)}\rangle={\mathcal
N}\sum_{n=0}^{d-1} \gamma_n|n\rangle,
\end{equation}
where the unknown input optical state $|\psi\rangle$  is truncated
to obtain the state $|\phi^{(d)}\rangle$, which is a finite
superposition of $d$ states (for a review see \cite{Adam01}). In
Eq. (\ref{N01}), $\mathcal{N}$ is the normalization constant,
which will be dropped from equations in the following, thus the
sign $\sim$ will be used instead of equality to denote that the
state should be normalized. This transformation is achieved by
conditional measurement and teleportation. This process was
originally described by Pegg, Phillips and Barnett to obtain a
superposition state of $d=2$ of the form $|\phi^{(2)}\rangle\sim
\gamma_0|0\rangle+\gamma_1|1\rangle$ by truncating a coherent
state $|\psi\rangle=|\alpha\rangle$, and was named as the {\em
quantum scissors device} (QSD) \cite{Pegg1,Pegg2}. Later we have
worked out a theoretical treatment of the QSD by proposing an
experimentally realizable scheme and discussing how arbitrary
superposition states of $d=2$ can be generated by this simple
scheme \cite{Sahin1,Sahin2}. The first experiment was performed by
Babichev {\em et al.} \cite{Babichev}. An extension of the
original QSD scheme to $d=3$ was proposed Koniorczyk {\em et al.}
by a simple modification of the original QSD scheme
\cite{Koniorczyk}. The original QSD scheme is an interesting one
because it finds its direct application as a basic element of
single-rail version of the linear-optical quantum computer.
Moreover, it is not only a truncation scheme but also a
communication scheme for superposition states of arbitrary $d$.

The drawback of the original QSD scheme is that it enables
generation of truncated states up to $d=3$. In this paper, we
extend the results of \cite{Adam05} to describe an application of
a modified version of the multiport Mach-Zehnder interferometer in
the configuration of Zeilinger {\em et al.} \cite{Zeilinger},
which has been experimentally demonstrated \cite{Reck,Mattle}. The
important difference between the original multiport interferometer
and the modified version discussed here is the elimination of the
apex BS so that the direct path from the input field to the output
field is eliminated. This is crucial for the truncation scheme as
we want the process to be done via teleportation.

In the following, we will introduce the generalized QSD scheme
based on multiport interferometer and give some examples of the
possible truncated states. Then we will discuss how the same
scheme can be used as a quantum punching device, which eliminates
selectively some Fock states from the original superposition state
and makes holes in the Fock-state expansion by proper choices of
conditional measurement and input states.

%------------------------------------------------------------------
\section{Multiport interferometer as quantum scissors device}

\begin{figure}
 \epsfxsize=8cm \epsfbox{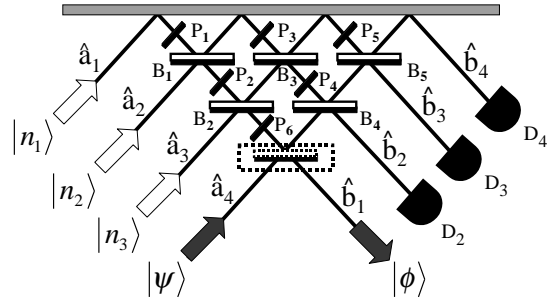}
\caption{Generalized eight-port quantum scissors device (QSD).
Notation: $|\psi\rangle$ is the input state to be truncated,
usually a coherent state $|\alpha\rangle$; $|n_j\rangle$ -- input
Fock states; $|\phi\rangle$ -- output state, selectively truncated
or punched; $D_j$ -- photon counters; $B_j$ -- beam splitters;
$P_j$ -- phase shifters; $\hat{a}_j$ and $\hat{b}_j$ -- input and
output annihilation operators, respectively. The beams reflected
from the white surface of the beam splitters are $\pi$ phase
shifted, while the reflections from the black surface are without
any phase shift. The beam splitter drawn with a dotted line and
bounded by the dotted box is the one to be removed to obtain the
QSD from the conventional multiport interferometer.} \label{Fig:1}
\end{figure}

A schematic diagram of the eight-port Mach-Zehnder interferometer
in the configuration of Zeilinger {\em et al.} and the generalized
QSD is given in Fig. \ref{Fig:1}. In a special case, the original
Pegg-Phillips-Barnett scheme of QSD can be considered as a
six-port interferometer presented in Fig. \ref{Fig:2}. The beam
splitter shown with the dotted lines corresponds to the apex BS
that is to be removed from the interferometer to obtain the
generalized QSD scheme. If we define the $N$-mode input state as
$|\Psi\rangle$ then the $N$-mode output state will be
$|\Phi\rangle=\hat{U}|\Psi\rangle$, where $\hat{U}$ is the unitary
operator describing the evolution of the input state in the
interferometer. Denoting the annihilation operators at the input
and output ports as column vectors ${\bf
\hat{a}}\equiv[\hat{a}_1;\hat{a}_2;\dots;\hat{a}_N]$ and ${\bf
\hat{b}}\equiv[\hat{b}_1;\hat{b}_2;\dots;\hat{b}_N]$,
respectively, we obtain ${\bf \hat{b}}=\hat{U}^\dag {\bf
\hat{a}}\hat{U}=S{\bf \hat{a}}$, where
${S}=P_6B_5P_5B_4P_4B_3P_3B_2P_2B_1P_1$ is the scattering matrix
obtained by multiplying the scattering matrices of the beam
splitters, $B_i$ and phase shifters $P_i$ used in the scheme from
the input to the output. We assume $B_i$ ($i=1,...,5$) to be
described by a real $2\times 2$ matrix $[t_i,r_i;-r_i,t_i]$
embedded in a $4\times 4$ matrix, where $t_i^2$ and $r_i^2$ being
the BS transmittance and reflectance, respectively. Internal phase
shifts of BSs can formally be included using external phase
shifters described by parameters $\xi_i$. For simplicity, we
analyze the system without $P_6$, i.e. assuming $\xi_6=0$.

Now, considering that at the input port we have the Fock state
$|\Psi\rangle=|n_1,\dots,n_N\rangle\equiv|{\bf n}\rangle$, the
output state is found as
\begin{equation}\label{N02}
|\Phi\rangle=\hat{U}|{\bf n}\rangle=\frac{1}{\sqrt{n_1!\cdots
n_N!}}\sum_{{\bf j}=1}^N \prod_{l=1}^\nu
S_{j_lx_l}\hat{a}^\dagger_{j_l}|{\bf 0}\rangle,
\end{equation}
where $S_{j_lx_l}$ are the elements of the unitary scattering
matrix $S$,  $\nu=\sum_i n_i$ is the total number of photons, and
$\sum_{{\bf j}}$ stands for the multiple sum over
$j_1,j_2,\dots,j_\nu$. Moreover, $x_l=j$ for $\sum_{i=1}^{j-1}n_i
<  l  \le \sum_{i=1}^{j}n_i$ and $j=1,...,N$.

%------------------------------------------------------------------
\section{Selective state truncations}

\subsection{Quantum scissors device}

In a truncation scheme, we are interested in obtaining a
superposition state by truncating an input optical state, which is
usually a coherent state. Therefore, in the generalized QSD
scheme, based on the multiport interferometer shown in Fig.
\ref{Fig:1}, we consider the state $|\psi\rangle$ as one of the
inputs. In that case, for the eight-port interferometer we can
write the total input state as
$|\Psi\rangle=|n_1\rangle_1|n_2\rangle_2|n_3\rangle_3|\psi\rangle_4$.
Now assume that the detectors at the output ports detect $N_2$,
$N_3$ and $N_4$ photons whose sum is the total number of photons
input into the interferometer, and satisfies the relation
$N_2+N_3+N_4=n_1+n_2+n_3=d-1$. This means that we project the
total output state $|\Phi\rangle_{1,2,3,4}$ onto the detected
states $|N_2\rangle_2|N_3\rangle_3|N_4\rangle_4$. Then the state
at the first output mode becomes
\begin{figure}
 \epsfxsize=8cm \epsfbox{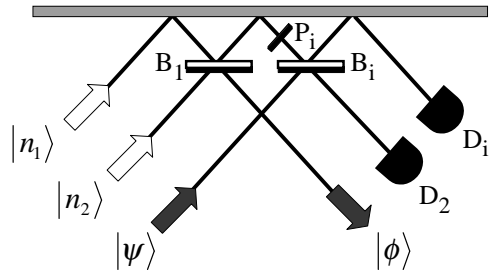}
\caption{Representation of the original Pegg-Phillips-Barnett
scheme of QSD as a six-port interferometer scheme. Notation is the
same as in Fig. \ref{Fig:1}, where $i=4$ (or equivalently $i=3$).}
\label{Fig:2}
\end{figure}
\begin{equation}\label{N03}
 |\phi\rangle\sim \;_2\langle N_2|\;_3
 \langle N_3|\;_4\langle N_4|\Phi\rangle=
 \sum_{n=0}^{d-1}c_n^{(d)}\gamma_n|n\rangle,
\end{equation}
where $c_n^{(d)}=\langle n ,N_2,N_3,N_4|\hat{U}|
n_1,n_2,n_3,n\rangle$ depends on the beam-splitter transmittances
${\bf T}\equiv[t_1^2,t_2^2,t_3^2,t_4^2,t_5^2]$ and phase shifts
${\mbf \xi}\equiv[\xi_1,\xi_2,\xi_3,\xi_4,\xi_5]$. Then our task
is to find ${\bf T}$ and ${\mbf \xi}$ according to the desired
state at the output in such a way that the fidelity of the output
state to the desired state is maximized.

It is seen from Figs. \ref{Fig:1} and \ref{Fig:2} that eliminating
the third modes at the input and output, and removing the
components on the path from the third input to the third output,
the eight-port interferometer becomes the original QSD (six-port)
when $|n_1\rangle_1|n_2\rangle_2=|1\rangle_1|0\rangle_2$ and
$|N_2\rangle_2|N_4\rangle_4=|1\rangle_2|0\rangle_4$. In this case,
the optimized solution with the highest probability of successful
truncation, that is corresponding to the output state
\begin{eqnarray}
 |\phi^{(2)}\rangle\sim \gamma_0|0\rangle+\gamma_1|1\rangle,
\label{N03a}
\end{eqnarray}
becomes ${\bf T}=[t_1^2=1/2,t_4^2=1/2]$ and ${\mbf
\xi}=[\xi_4=\pi]$. In the same way Koniorczyk's QSD is obtained in
the same six-port interferometer with
$|n_1\rangle_1|n_2\rangle_2=|1\rangle_1|1\rangle_2$ and
$|N_2\rangle_2|N_4\rangle_4=|1\rangle_2|1\rangle_4$. Then we find
that there are four solutions for the successful truncation with
the highest probability to obtain the state
\begin{eqnarray}
  |\phi^{(3)}\rangle\sim
\gamma_0|0\rangle+\gamma_1|1\rangle+\gamma_2|2\rangle.
\label{N03b}
\end{eqnarray}
These solutions are ${\bf T}_1=[t_1^2=t_4^2=(3-\sqrt{3})/6]$,
${\bf T}_2=[t_1^2=t_4^2=(3+\sqrt{3})/6]$ if ${\mbf \xi}
=[\xi_4=0]$, and ${\bf T}_3=[t_1^2=(3-\sqrt{3})/6,
t_4^2=(3+\sqrt{3})/6]$, ${\bf T}_4=[t_1^2(3+\sqrt{3})/6,
t_4^2=(3-\sqrt{3})/6]$ if ${\mbf \xi}=[\xi_4=\pi]$. The first two
solutions were given by Koniorczyk {\em et al.} \cite{Koniorczyk},
but the rest have been found by us.

For the generalized QSD with the modified eight-port
interferometer, we are more interested in the device to act as a
QSD with a simple solution than the optimality of the solutions.
We find that an input coherent state at the fourth-mode of the
input can be truncated to give the output state
\begin{eqnarray}
 |\phi^{(4)}\rangle\sim
\gamma_0|0\rangle+\gamma_1|1\rangle+\gamma_2|2\rangle+\gamma_3|3\rangle
\label{N03c}
\end{eqnarray}
by inputting single-photon states at
$|n_1\rangle_1|n_2\rangle_2|n_3\rangle_3
=|1\rangle_1|1\rangle_2|1\rangle_3$ and by the conditional
measurement $N_2=N_3=N_4=1$. We find a number of solutions for
transmittances and phase shifts in the QSD, for which the input
state is truncated to form (\ref{N03c}). One simple solution is
given by ${\bf T}=[1/3,1/4,1,1/3,1/2]$ with ${\mbf
\xi}=[0,0,0,0,\pi/2]$.

Consequently, various output states with desired coefficients can
be obtained by proper choices of ${\bf T}$ and ${\mbf \xi}$
provided that the total number of photons detected at the output
detectors equal to the total number of input photons. For example,
by inputting
$|n_1\rangle_1|n_2\rangle_2|n_3\rangle_3=|1\rangle_1|2\rangle_2|1\rangle_3$
and detecting
$|N_2\rangle_2|N_3\rangle_3|N_4\rangle_4=|1\rangle_2|2\rangle_3|1\rangle_4$,
a truncated output state in the form
\begin{eqnarray}
  |\phi^{(5)}\rangle\sim
\gamma_0|0\rangle+\gamma_1|1\rangle+\gamma_2|2\rangle
+\gamma_3|3\rangle+\gamma_4|4\rangle
\label{N1}
\end{eqnarray}
can be obtained by choosing the BS and PS parameters as ${\bf
T}=[0.305,0.388,1,0.817,0.184]$ with ${\mbf \xi}=[0,0,0,\pi,0]$.
For larger dimensional output states, it is difficult to obtain
analytical solutions, therefore solutions are found by numerical
analysis on condition that the fidelity of the output state to the
desired one is the highest.

%------------------------------------------------------------------
\subsection{Quantum punching device}

Here, we consider the cases where the output state obtained by
truncating the input state $|\psi\rangle$ has some of its Fock
states removed. Let us assume that $|k_1\rangle,|k_2\rangle,...$
are removed, then the output state is written as
\begin{equation}\label{N06}
|\phi^{(d)}_{{\rm punch}}\rangle\sim \sum_{n=0\atop{n\neq
k_1,k_2,...}}^{d-1}\gamma_n|n\rangle.
\end{equation}
We call this kind of process, which makes holes in the Fock-state
expansion, the quantum punching device (QPD). We have observed
that by choosing proper BSs and PSs we can achieve this kind of
state engineering using the multiport interferometer. For example
in the state $|\phi^{(4)}\rangle$, given by (\ref{N03c}), we can
punch out (or remove) the state $|2\rangle$ to get
\begin{eqnarray}
  |\phi^{(4)}_{{\rm punch}~01\bullet 3}\rangle\sim
\gamma_0|0\rangle+\gamma_1|1\rangle+\gamma_3|3\rangle,
\label{N03e}
\end{eqnarray}
by choosing ${\bf T}=[(7+\sqrt{21})/14,1/3,1,1/2,(5-\sqrt{5})/10]$
with ${\mbf \xi}={\bf 0}$ (i.e., with all phase shifts equal to
zero). Hereafter, we assume the input Fock states
$|n_i\rangle_i=|1\rangle_i$ ($i=1,2,3$) and all measurement
outcomes equal to one. In the same way, we can obtain the state
\begin{eqnarray}
|\phi^{(4)}_{{\rm punch}~\bullet123}\rangle\sim
\gamma_1|1\rangle+\gamma_2|2\rangle+\gamma_3|3\rangle \label{N03f}
\end{eqnarray}
for ${\bf T}=[(7+\sqrt{21})/14,1/3,1,1/2,(2-\sqrt{2})/4]$ with
${\mbf \xi}={\bf 0}$.  We have observed that superpositions of any
two Fock states
\begin{equation}\label{N07}
|\phi^{(4)}_{{\rm
punch}~kl\bullet\bullet}\rangle\sim\gamma_k|k\rangle+\gamma_l|l\rangle
\end{equation}
can be obtained as special cases of the truncation process, e.g.,
for ${\mbf \xi}={\bf 0}$ and the transmittances given by
\begin{eqnarray}
\label{N08}
 |\phi^{(4)}_{{\rm punch~0\bullet2\bullet}}
 \rangle&{\rm for}& {\bf T}=[1,\frac12,1,1,\frac12],\nonumber\\
 |\phi^{(4)}_{{\rm punch~\bullet1\bullet3}}\rangle&{\rm for}& {\bf
 T}=[\frac12,T,1,T,\frac12],\\
|\phi^{(4)}_{{\rm punch}\, 0\bullet\bullet3} \rangle&{\rm for}&
{\bf T} =[T',\frac{1}{2},1, \frac{1}{6},T''],\nonumber
\end{eqnarray}
where $T=(3-\sqrt{3})/3$, $T'=(1-\sqrt{5/133})/2$, and
$T''=(1+3\sqrt{3/155})/2$. It is interesting to see that one can
synthesize two and three photon Fock states in the
$|\phi^{(4)}\rangle$ process by choosing ${\bf
T}=[1,1/2,1/3,1/2,1]$ and ${\bf T}=[1/2,1/2,1,1/2,1/2]$,
respectively, and assuming ${\mbf \xi}={\bf 0}$ except
$\xi_5=\pi/2$ in the latter case.

It must be noted we have given only some specific examples, which
guarantee the desired output state but the solutions are usually
not optimized for the success probability.

%------------------------------------------------------------------
\section{Conclusion}

We have shown that the original Pegg-Phillips-Barnett scheme of
QSD can be generalized using multiport Mach-Zehnder
interferometers in the configuration of Zeilinger {\em et al.}
\cite{Zeilinger}. The original QSD scheme can be represented as a
six-port interferometer. The multiport interferometer approach can
help us not only to truncate a coherent state to obtain a
superposition state up to an arbitrary Fock state but it also
enables selective truncation of a given state and selective
removal of Fock-state components from it. As it was the case in
the original QSD, the generalized one also produces the desired
output state with very high fidelity when the input state to be
truncated is a weak coherent state. A hard problem we face in this
scheme is the optimization of the solutions to obtain the highest
probability of truncation when $d$ is high. In the present study,
we did not focus in optimizing our solutions but in showing that
the scheme is working as a truncation or punching device. The
effects of imperfections (such as non-ideal photon counting and
non-ideal single-photon source) in the scheme are currently being
investigated.

\begin{acknowledgments}
AM was supported by the Polish State Committee for Scientific
Research (grant No. 1 P03B 064 28). JB was supported by the Czech
Ministry of Education (grant MSM6198959213).

\end{acknowledgments}

\end{document}